%% file: main.tex
\documentclass[preprint,journal,hideappendix]{vgtc}        
\usepackage{balance}
\usepackage[utf8]{inputenc}

\onlineid{1581}

\vgtccategory{Research}
\vgtcpapertype{Theoretical \& Empirical}
\title{CatPAL: Task-Aware Learning for Categorical Palette Recommendation}

\author{%
  Chin Tseng,
  Arran Zeyu Wang,
  Yunqi Li, and 
  Danielle Albers Szafir
}

\authorfooter{
  \item
  	Chin Tseng, Arran Zeyu Wang, Yunqi Li, and Danielle Albers Szafir are with the University of North Carolina at Chapel Hill (UNC).
  	E-mail: {chint, zeyuwang, yunqili, danielle.szafir}@cs.unc.edu
}

\abstract{Designing effective categorical palettes 
requires balancing a range of factors, including perceptual
distinctiveness, category count, and 
task effectiveness. The effectiveness of categorical encodings can vary substantially depending on the target analytical tasks; however, existing recommendation tools largely ignore task context when evaluating palette quality, resulting in inconsistent performance across tasks.
\fix{We synthesize} findings from \fix{a series of} multi-stage user studies into a unified model of task-based effectiveness for color encodings, shape encodings, and their redundant combination across category counts and seven common scatterplot tasks. Our results show that \emph{task and palette choice jointly influence perceptual accuracy}: different color and shape palettes exhibit varying levels of robustness across tasks, indicating that palette effectiveness is 
task-dependent.
We estimate task-specific perceptual strengths for 39 colors and 39 shapes using Bradley-Terry models, refined through adaptive sampling to target uncertain and task-sensitive comparisons. We further quantify cross-channel interactions using a \emph{redundant gain} ($\Delta G$) metric
to model performance across color and shape pairings.
We then train a predictive model that scores candidate palettes based on task, category count, and perceptual features. 
This model drives effective palette recommendations in 
\href{https://catpal-palette-automation.web.app/}{\emph{CatPAL}}, \fix{a task-aware palette recommendation system grounded in empirical data}
responsive to user constraints. 
Our findings highlight the importance of selecting categorical palettes aligned with specific analytical tasks and demonstrate how task-aware modeling enables more reliable palette design. \emph{CatPAL} translates empirical results into a practical tool that supports user-specified colors or shapes and returns ranked palette recommendations adaptable to a range of tasks.
}

\keywords{Categorical perception, shape perception, multiclass scatterplots, visualization effectiveness, quantitative study}

\teaser{
  \centering
\vspace{-0.5em}
  \includegraphics[width=\linewidth]{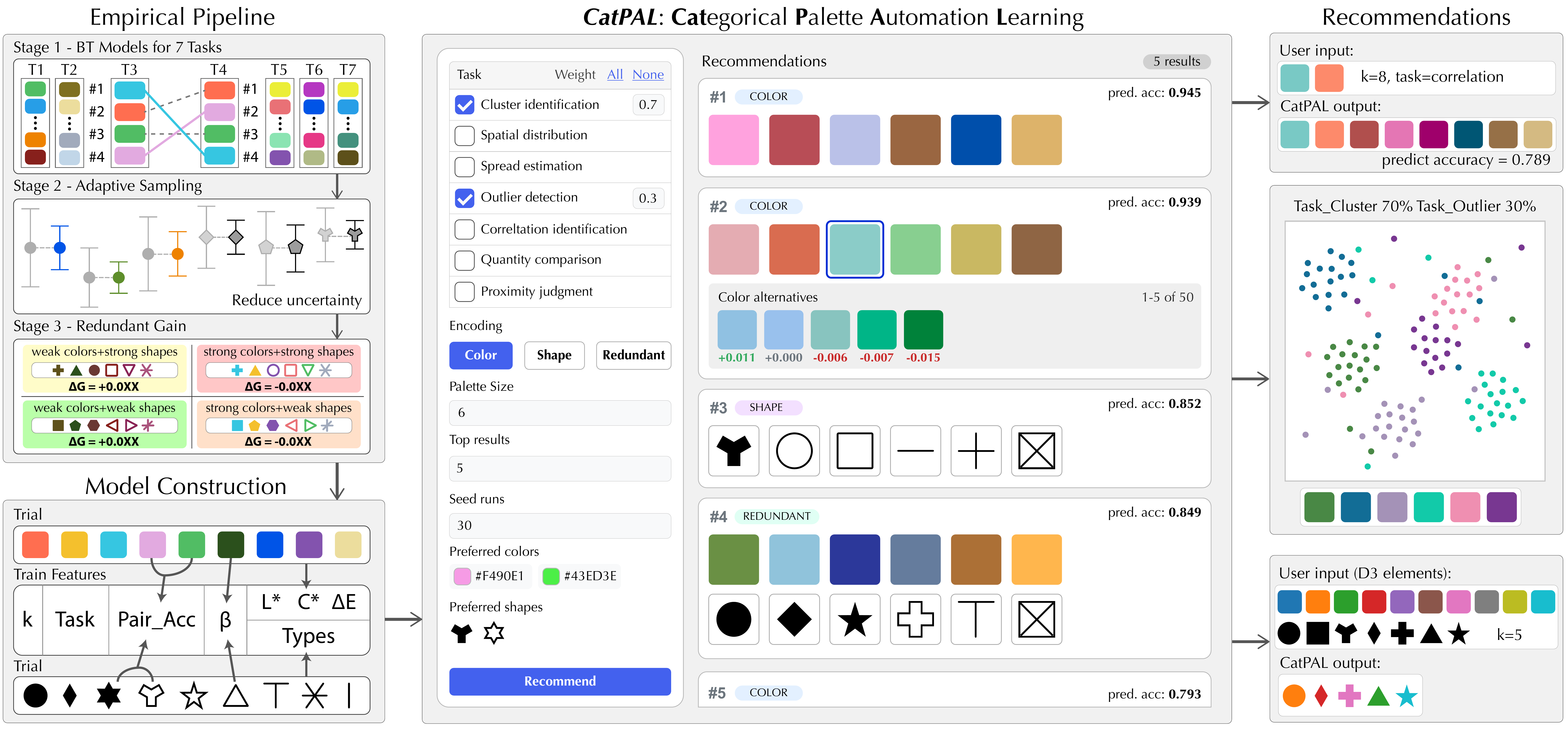}
\vspace{-1em}
  \caption{\textbf{\textsc{CatPAL}: A task-aware categorical palette recommendation system for scatterplots.} Given a set of scatterplot tasks and a category number, \emph{CatPAL} selects top-performing palettes that maximize perceptual accuracy. Different tasks lead to different palette performance, as color and shape encodings vary in effectiveness across tasks. \fix{Drawing on empirical data,} \emph{CatPAL} captures these task-dependent effects and recommends palettes optimized for the task across color, shape, and redundant encodings. Users can provide preferred colors or shapes and interactively refine results, enabling flexible palette design.}
  \label{fig:teaser}
\vspace{-0.5em}
}

\graphicspath{{figs/}{figures/}{pictures/}{images/}{./}} 

\usepackage{tabu}                      
\usepackage{booktabs}                  
\usepackage{lipsum}                    
\usepackage{mwe}                       
\usepackage{ccicons}                   

\usepackage{mathptmx}                  
\usepackage{amsmath}

\newcommand{\fix}[1]{\textcolor{black}{#1}}

\begin{document}

\renewcommand{\sectionautorefname}{Section}
\renewcommand{\subsectionautorefname}{Section}
\renewcommand{\subsubsectionautorefname}{Section}

\maketitle

\section{Introduction}
\input{sections/sec-01-intro}

\section{Background and Related Work}

\input{sections/sec-02-rw}

\section{Methodological Framework}
\label{sec-pipeline}
\input{sections/sec-03-method}

\section{Experiment One \& Two: Task-Aware Perceptual Modeling with Adaptive Sampling}
\label{sec-exp1-exp2}
\input{sections/sec-04-BTmodel}

\section{Experiment Three: Gain Study on Redundancy}
\label{sec-exp3}
\input{sections/sec-05-redundant}

\section{Task-Aware Categorical Model}
\label{sec-model}
\input{sections/sec-06-model}

\section{Discussion}
\input{sections/sec-07-discussion}

\section{Conclusion}
\input{sections/sec-08-conclusion}

\acknowledgments{
We thank the reviewers for their insightful comments.
This work was supported by the National Science Foundation under grant NSF IIS-2046725 and NSF IIS-1764089.
}

\balance
\bibliographystyle{abbrv-doi-hyperref-narrow}
\bibliography{main}


\end{document}

%% file: sections/sec-01-intro.tex
Visualizations often must convey patterns in categorical data, such as group comparisons in exploratory scatterplots~\cite{satyanarayan2017vegalite} and class boundaries in 
multidimensional embeddings~\cite{smilkov2016projector}.
However, assembling effective categorical palettes requires considering a range of factors, including both objective (e.g., the encoding channel \cite{tseng2026redundant} and data distribution \cite{zimnicki2023effects,lu2020palettailor}) and subjective (e.g., affect \cite{braun2025affective} and aesthetics \cite{gramazio2016colorgorical}) factors.
The complexity of palette design makes it challenging to create palettes without substantial prior expertise. 
Yet palette design remains a critical element of effective visualization. Well-chosen palettes facilitate rapid and accurate interpretation, while poorly designed palettes
can obscure patterns, mislead viewers, and render a graphic ineffective~\cite{ware2012information, zhou2015survey}.

Existing tools for palette selection 
\fix{aim to make} effective visualization design more accessible, primarily by optimizing for perceptual metrics~\cite{gramazio2016colorgorical} to ensure colors are discriminable~\cite{sharma2005ciede2000} or by 
integrating aesthetic principles~\cite{schloss2011aesthetic}.
However, these tools largely focus on designing categorical colors, failing to consider alternative encodings like shape. They also lack the means for considering the context of a visualization's data and use into design. 
For example, the number of visualized categories can affect the robustness of a palette \cite{tseng2023evaluating}. 
Task can influence the effectiveness of continuous color encodings \cite{tominski2008task,dasgupta2018effect} and other visual encoding choices \cite{albers2014task,saket2018task} and may have similar effects on categorical palettes. 
The perceptual demands of different tasks vary~\cite{shneiderman2003eyes}, meaning a palette optimized for one task may be suboptimal for another.
Furthermore, the common practice of using redundant encodings unifying shape and color~\cite{tseng2026redundant, nothelfer2017redundant}---a strategy often assumed to be universally beneficial---has not been systematically \fix{evaluated} across different task contexts. 
This paper addresses these gaps in categorical palette design by introducing a task-aware empirical model of categorical palette 
design that jointly considers encoding channel (color, shape, and their redundant combination), task context, and category count to support principled palette recommendation.

\fix{We modeled palette effectiveness through} a series of crowdsourced experiments on Amazon Mechanical Turk (MTurk), measuring people's ability to reason about categories encoded with color and shape across 
seven scatterplot tasks (cluster identification, spatial localization, spread, outlier detection, correlation, numerosity, and proximity estimation) and three category sizes ($k = 3, 6, 9$).
The task selection covers different facets of established scatterplot task taxonomies~\cite{sarikaya2017scatterplots} and 
reflects analytic demands typical in real-world use~\cite{wilkinson2005graph}.
Our experiments form a three-stage adaptive data collection pipeline, where the collected empirical data is used to 
generate a model covering a broad range of colors and shapes for different tasks and category demands. 
We start with an anchoring stage to estimate per-task Bradley--Terry (BT) strength scores~\cite{hunter2004mm} for a representative set of colors and shapes. We then adaptively sample new combinations, focusing on color and shape pairs where the model was least confident to refine those estimates. Finally, we consider cross-encoding palette designs by measuring redundancy gain ($\Delta G$) 
across color--shape combinations drawn from 
the BT rankings.
We use the data from this empirical pipeline train task-oriented models that predict how accurately people can 
complete a target task for a given palette and category number.

\fix{Our experiments demonstrate that good palettes are not simply distinguishable but must consider holistic performance for the target use context. }
Comparing against established palettes (e.g., Tableau~\cite{tableau}, ColorBrewer~\cite{harrower2003colorbrewer}), tool-generated palettes (e.g., Colorgorical~\cite{gramazio2016colorgorical}, IWantHue~\cite{iwanthue}), and palettes produced by large language models shows that our model-optimized palettes consistently achieve higher predicted accuracy across tasks and category sizes. This suggests that task context and empirical grounding are key factors that existing palette design approaches do not fully capture, and that simple perceptual cues (e.g., $\Delta E$ or other color statistics) alone are insufficient to explain performance. 

We instantiate our empirical models in \textbf{CatPAL} (\textbf{Cat}egorical \textbf{P}alette \textbf{A}utomation \textbf{L}earning), a web-based tool that takes a user's target tasks, preferred encoding channel, preferred colors or shapes, and/or category numbers as input and generates color, shape, or redundant palettes 
tailored to the specified design constraints. 
Users can interactively refine recommendations 
using alternatives suggested by CatPAL to make informed trade-offs between optimality and personal or aesthetic preference.
Unlike existing tools that optimize color in isolation, CatPAL jointly recommends color, shape, and redundant encodings based on the user’s task goals, data size, and other preferences, translating empirical findings into a practical and accessible tool to democratize the process of creating effective categorical palettes. 

Our contributions include:

    \noindent \emph{A Large-Scale Empirical Study} of categorical encoding (color, shape, and color+shape) discriminability across seven scatterplot tasks and three category sizes ($k = 3, 6, 9$). We use a novel three-stage adaptive crowdsourced pipeline to collect pairwise accuracy data over a broad set of colors and shapes and fit per-task Bradley–Terry models to understand how encoding effectiveness varies across encodings as a function of task and category count.
    
    \noindent \emph{A Task-Aware Predictive Model} that estimates palette 
    effectiveness from encoding features and task context, and enables 
    insight into when and how redundant color–shape encodings provide additional benefit.
    
    \noindent \emph{An Interactive Palette Design Approach} embodied in CatPAL, a 
    web-based palette recommendation tool that operationalizes the model, allowing users to specify task and category count to interactively generate and refine color, shape, or redundant categorical palettes.

%% file: sections/sec-02-rw.tex

\subsection{Palette Design for Data Visualization}

Visualization research has long focused on developing effective color encodings.
Foundational work by Ware~\cite{ware1988color} and Healey~\cite{healey1996choosing} established principles for using color in information displays, 
emphasizing the importance of 
perceptual color characteristics in categorical encoding. 
These principles led to the development of seminal tools like ColorBrewer~\cite{harrower2003colorbrewer} which provided cartographers and designers with 
perceptually-based color schemes for maps.
Subsequent tools have 
incorporated additional design considerations and automated techniques for designing color encodings \cite{hong2024cieran} (see Zhou \& Hansen \cite{zhou2015survey} for a survey). 
For example, Colorgorical~\cite{gramazio2016colorgorical} uses a weighted combination of perceptual distance, name difference/uniqueness, and pairwise aesthetic preference to generate distinguishable palettes.
ColorCrafting~\cite{smart2019color} models the underlying geometric structures of expert colormaps and applies these models to target colors to generate 
sequential color encodings.
Color Maker~\cite{salvi2024color} 
allows people to integrate accessibility considerations into colormap design.
Palettailor \cite{lu2020palettailor} integrates spatial considerations into selecting and assigning color palettes to visualizations. 

While existing tools ensure discriminability and aesthetic quality, their core optimization is typically based on general-purpose perceptual principles or style transfer, neglecting other factors such as 
analytical tasks \cite{tominski2008task} or data distributions \cite{zimnicki2023effects,tseng2023evaluating}. 
Factors like color area, background contrast, and hue separation all impact palette performance~\cite{smallman1990segregation}.
A range of experiments offer models for many of these factors, like 
mark size and shape~\cite{szafir2014adapting, szafir2018modeling,smart2019color}, aesthetics and semantics~\cite{schloss2010aesthetics, schloss2018color, schloss2018mapping, schloss2024color}, affect \cite{braun2025affective}, and specific colormap structures such as rainbows~\cite{liu2018somewhere,ware2018measuring,reda2020rainbows}, categorical palettes~\cite{tseng2023evaluating}, and continuous colormaps~\cite{tseng2024revisiting}.

While color encodings have received significant attention in palette design tools, relatively less emphasis has been placed on alternative encodings like shape. However, shape palettes also vary in their effectiveness \cite{tseng2024shape}, with the set of factors influencing shape encoding design being significantly less well defined \cite{burlinson2017open,tseng2024shape,demiralp2014learning}. Interactions between color and shape channels, especially in redundant encodings where both channels represent the same data \cite{nothelfer2017redundant}, may also influence the efficiency of encoding palettes for categorical data \cite{smart2019color,tseng2026redundant}.
Fewer tools also support people in designing such palettes. Notably, Shape It Up~\cite{tseng2024shape} supports shape palette design specifically, while CatPAW~\cite{tseng2026redundant} 
enables redundant palette design. These tools both draw from empirical data to generate palettes using weighted statistical scoring functions; however, they cannot account for critical design factors like target task and do not enable significant iteration in their designs.

\subsection{Categorical Perception}


The perceptual system's ability to rapidly and accurately group visual elements is fundamental to reading visualizations~\cite{goldstone2010categorical}.
Categorical visualization takes advantage of this ability to help communicate categorical data, such as different teams in sports visualization or organisms in biology. 
Much of the research in categorical perception has focused on color palettes. For example, 
Heer \& Stone \cite{heer2012color} modeled studied how people name colors, providing a basis for selecting \emph{nameable} colors. 
Subsequent work used color names to drive palette design and improve search and communication in categorical data \cite{lin2013selecting,setlur2016linguistic}.
Color concept associations can help characterize palette effectiveness by mapping colors logically to their corresponding semantic concepts \cite{mukherjee2021context}. 

Other studies investigate the role of shape and other channels in representing categories. For example, perceptual kernels summarize subjective perceptual distances between predefined sets of shapes, colors, and sizes \cite{demiralp2014learning}. Burlinson et al. \cite{burlinson2017open} measured how different shape types (open versus closed shapes) influence category judgments. Gleicher et al. \cite{gleicher2013perception} explored how different encoding channels affect position estimation in multiclass scatterplots. More recent efforts have generalized this idea to exploring both color \cite{tseng2023evaluating} and shape \cite{tseng2024shape} palettes across different numbers of categories. Wang et al. \cite{wang2025characterizing} replicated these studies to demonstrate the potential role of subitizing and 
the cognitive limits of the human visual system~\cite{haroz2012capacity} by assessing performance using tasks varying in their underlying perceptual processes. However, this study focused on the broader relationship between category count and performance rather than specific palette design factors.
Our work extends this line of inquiry by systematically measuring categorical perception 
in tasks of varying complexity, from simple count estimation to complex cluster and sparsity detection.
This paradigm allows us to provide a more nuanced model of categorical data perception.

\subsection{Task-based Visualization Effectiveness}

The principle that visual design should be guided by the user's target task is 
central to visualization research~\cite{shneiderman2003eyes}.
A broad body of research characterizes the effectiveness of visualizations for different tasks
such as averages \cite{gleicher2013perception,correll2012comparing,hong2021weighted}, correlation \cite{rensink2010perception,harrison2014ranking,kay2015beyond}, and clustering~\cite{wilkinson2005graph,wang2019improving,sedlmair2012taxonomy}
(see Quadri \& Rosen for a survey~\cite{quadri2021survey}).
Visualization research has offered several task taxonomies that summarize the critical operations people perform with visualizations \cite{amar2005low,brehmer2013multi,schulz2013design}.
For example, Sarikaya et al.~\cite{sarikaya2017scatterplots} categorized 
a taxonomy of tasks for scatterplots that summarize twelve high-level types of tasks people use scatterplots to achieve. 

However, the interaction between categorical palette design choices and tasks remains unclear. 
Tominski et al. \cite{tominski2008task} characterize how continuous colormaps can be designed to support a range of tasks. ColorMoves \cite{samsel2018colormoves} allows scientists to assemble colormaps that highlight different ranges of the data. 
However, these approaches focus on continuous data and heuristics for inferring task. We lack grounded empirical models that can guide design. 
Our study directly addresses this limitation by creating a task-conditioned model of palette quality.
We quantify how the performance ranking of colors and shapes changes across tasks 
to provide a critical link between abstract task taxonomies and concrete design guidelines.

%% file: sections/sec-03-method.tex
\subsection{Experimental Pipeline}
\label{sec:pipeline}

Building a task-aware palette recommendation system requires 
understanding how effectively people estimate different statistics 
for every candidate color and shape as well as across multiple tasks and category counts. Specifically, our goal is not simply to rank 
preconstructed color or shape palettes, but to learn what encoding channel (color, shape, or redundant) to use and assemble palettes from the set of available colors and shapes to 
support target tasks for
a given category number.
We need empirical data to generate models of task performance across these factors.
However, 
the overall problem space is larger than a single experiment can address directly: experiments must determine
which colors or shapes to evaluate, how many categories to use, 
the set of target tasks being performed, and whether the palettes 
use color, shape, or both. Moreover, we aim to maximize the generalizability and usability of the recommendation system by developing a model that can support a wide range of colors and 
account for user preferences.

To manage both our system goals and the practical limitations of such a large design space, we
designed a 3-stage data collection pipeline in which the data collected in each phase addresses a different part of the recommendation problem while enabling study design at a practical scale.
We use a multistage pipeline as 
exhaustive coverage is infeasible: testing all $\binom{39}{2} = 741$ pairwise combinations from our candidate encoding pool per encoding (color, shape), across seven tasks and category counts ($k=2-10$), yields $741 \times 2 \times 7 \times 9 = 93{,}366$ conditions. Uniform sampling across this space would leave too few observations per condition to support reliable estimation. To address sample sparsity, we augment a theoretically-grounded sampling approach (Stage 1) with an adaptive sampling approach based on the results of the initial sampling to focus data on more uncertain areas of the design space (Stage~2). 
Second, palette rankings may not be consistent across different analytic tasks. 
We can leverage adaptive sampling (Stage~2) to prioritize color or shape pairs with higher performance variation across tasks to derive separate models for different tasks. 
Third, single-channel estimates alone are insufficient for predicting effective redundant encodings~\cite{tseng2026redundant}. Measuring the interaction exhaustively across all 
possible color--shape combinations would again be impractical.
We therefore use the results from the first two stages to select relevant color and shape palettes 
for
targeted data collection for redundancy gains
across tasks (Stage 3). 

\autoref{fig:teaser} summarizes the full pipeline.
\textbf{Stage~1} establishes an initial scaffold over the item (color, shape) space. 
As described in \autoref{sec-exp1-exp2}, we select a series of color and shape pairs from past studies to collect data about performance across a series of tasks (see \autoref{tab:tasks}). We then collect data about performance of palettes using these pairs to construct a series of Bradley--Terry (BT) models summarizing the resulting performance space. BT models assign a latent strength score to each encoding item (i.e., every color or shape) by fitting a log-odds model over observed correct/incorrect outcomes from pairwise comparisons, creating a continuous ranking of colors and shapes by their performance for each task and category count.
We use Bradley--Terry because it is well-suited to pairwise comparison data, allowing us to reasonably model encoding pairs, and remains effective under sparse, unbalanced observations, allowing us to estimate latent item strengths without exhaustively sampling every possible pair.
The models generated in this stage provide a stable initialization: they give us preliminary estimates of per-task and per-k pairwise encoding strengths, and additionally let us compute uncertain areas within the encoding space that 
help identify where additional data are most needed.

\textbf{Stage~2} uses the Stage~1 estimates to expand coverage 
by selecting encoding pairs that offer significant insight for the model. 
We score residual pairs from the overall corpus sampled in Stage 1 based on how much they improve the model. Scores are based on whether the pairs fill gaps left by the anchor design from Stage~1, whether they
are situated near encoding pairs of high variance across tasks, and whether prior evidence suggests they are 
uncertain (see \autoref{sec-exp1-exp2}). We then collect data using the highest-scoring encoding pairs
and re-fit the BT models on the combined Stage~1+2 dataset.
These models produce the final task-specific single-channel strength scores $\beta_i^t$
and 
identify 
which colors and shapes are strong or weak for each task and category count.

\textbf{Stage~3} 
then uses the results of the Stage 1 and 2 models to systematically characterize efficiency of redundant encodings (see \autoref{sec-exp3}).
We first categorize colors and shapes into four groups using the Stage~1+2 BT estimates. We classify encoding pairs
into strong and weak performing sets, then pair them in a $2 \times 2$ quadrant design (strong--strong, strong--weak, weak--strong, weak--weak).
Using both strong and weak performing encoding pairs allows us to systematically vary the baseline quality of each channel and examine how redundancy interacts with single-channel strength. By spanning all four quadrants, we introduce the variation needed for the model to learn when redundancy provides meaningful gains, when it offers little additional benefit, and how those gains depend on the strength of the individual channels.
We collect data using redundant pairs sampled across these four sets to estimate the \emph{redundancy gain}, $\Delta G = \mathrm{Acc}(color{+}shape) - \max(\mathrm{Acc}(color), \mathrm{Acc}(shape))$,
to model performance gains from redundancy.

\textbf{Model training} converts the empirical findings from all three stages into a 
recommendation engine (see \autoref{sec-model}). The final model ingests features derived from the Stage~1+2 BT strength estimates, pairwise accuracy, color parameters, shape types, tasks, category numbers, and the Stage~3 redundant-gain measurements. We train a predictive palette scorer using logistic regression models that estimate expected accuracy for candidate color, shape, or redundant palettes under user-specified task conditions and category number. We integrated the model into a web-based recommendation tool, CatPAL, that enables users to interactively create efficient categorical palettes with their own preferences.

\subsection{Task Choices and Justification}

Our data collection experiments use 
seven scatterplot analysis tasks to capture a diverse range of common analytical goals in multiclass data visualization. We derive these tasks from
Sarikaya et al.'s task taxonomy for scatterplots~\cite{sarikaya2017scatterplots}, with at least one task per 
distribution characterization (e.g., identifying clusters and spatial spread), data exploration, anomaly detection, correlation assessment, numerosity comparison, and distance judgment (\autoref{tab:tasks}). 


\begin{table}[t]
  \centering
  \caption{The seven analytical tasks used in our studies spanned
  a range of analytical goals~\cite{sarikaya2017scatterplots}: distribution characterization ($T_{cluster}$, $T_{localization}$, $T_{spread}$), anomaly detection ($T_{outliers}$),
    correlation assessment ($T_{correlated}$), numerosity comparison ($T_{points}$), and distance judgment ($T_{closest}$).}
    \vspace{-0.5em}
  \label{tab:tasks}
  \small
  \begin{tabular}{@{}clp{5.8cm}@{}}
    \toprule
    \textbf{\#} & \textbf{Task} & \textbf{Question} \\
    \midrule
    1 & Tightest cluster   & Which category forms the tightest cluster? \\[3pt]
    2 & Spatial localization        & Which category appears in only one half (top or bottom)? \\[3pt]
    3 & Most spread        & Which category is the most spread out? \\[3pt]
    4 & Most outliers      & Which category has the most outliers? \\[3pt]
    5 & Most correlated    & Which category is the most correlated? \\[3pt]
    6 & Most points        & Which category has the most points? \\[3pt]
    7 & Closest to a Reference       & Which category is closest to X? \\
    \bottomrule
  \end{tabular}
  \vspace{-2em}
\end{table}

We selected tasks reflecting aspects of the scatterplot task taxonomy that 
likely vary along the employed perceptual mechanisms, cognitive operations, and visual search strategies.
We aim to ensure that palette effectiveness is evaluated across different perceptual demands rather than variations of a single task type as palettes are characterized by their variation in individual visual features (e.g., color hue or lightness) and different perceptual mechanisms may process these features differently \cite{szafir2023visualization, elliott2020design}.
Visualization effectiveness is strongly task-dependent: different tasks rely on different perceptual processes, such as grouping~\cite{gramazio2014relation}, estimation~\cite{ware2012information}, comparison~\cite{gleicher2011visual}, or search~\cite{haroz2012capacity}.
Our selected tasks include:

\noindent \textbf{Tightest Cluster} ($T_{cluster}$):
Detecting and analyzing clusters 
is a fundamental operation in scatterplot analysis~\cite{sedlmair2012taxonomy}.

\noindent \textbf{Spatial Localization} ($T_{localization}$):
The spatial localization task reflects spatial localization and region-based browsing~\cite{sarikaya2017scatterplots}.


\noindent \textbf{Largest Spread} ($T_{spread}$):
The largest spread task captures dispersion patterns, asking people to identify the category with the greatest spatial spread (quantified as the largest convex hull)~\cite{wilkinson2005graph, cleveland1984graphical}. 

\noindent \textbf{Most Outliers} ($T_{outliers}$):
Scatterplots are often used for anomaly detection, 
where people find points that violate the general distribution of a cluster or category~\cite{wilkinson2005graph, wang2019improving}. 

\noindent \textbf{Most Correlated} ($T_{correlated}$):
Correlation is one of the most studied tasks in visualization perception research~\cite{harrison2014ranking, rensink2010perception, yang2018correlation}.

\noindent \textbf{Most Points} ($T_{points}$):
People often engage in numerosity estimation and counting
with scatterplots to understand variance in quantities between categories~\cite{wang2025characterizing}.

\noindent \textbf{Closest to a Reference} ($T_{closest}$)
The reference task represents distance comparison and relational spatial reasoning, asking people to 
compare distances \textit{between} categories~\cite{gleicher2011visual, gogolou2018comparing}. 

Please see Supplement for a full task justification for each task. 

Compared to prior work that focuses on a single task (e.g., correlation judgment~\cite{tseng2026redundant}), our task set enables us to examine how the perceptual effectiveness of categorical encodings varies across analytical contexts. This design allows us to systematically evaluate whether different color and shape palettes exhibit varying levels of robustness across tasks and construct
task-aware palette recommendations.

%% file: sections/sec-04-BTmodel.tex
Experiments~1 and~2 correspond to Stages~1 and~2 of our empirical pipeline (\autoref{sec-pipeline}), forming a two-phase approach to estimating per-color and per-shape perceptual strength 
across tasks.
Experiment~1 establishes an initial BT model using a set of anchor items whose mean accuracies across different performance varies in legacy studies ~\cite{tseng2024shape, tseng2026redundant}, providing preliminary estimates of per-task item strength and uncertainty. Experiment~2  uses those Stage~1 estimates to selectively expand coverage toward items with the highest variance across tasks, then refits the BT models on the combined Stage~1+2 dataset. Together, they produce the final task-specific BT strength and pairwise accuracy metrics for a pool of 39 colors and shapes derived from prior work---the core perceptual features used to score and rank candidate palettes in the CatPAL recommendation engine.

\subsection{Experimental Design}
Our approach is grounded in two hypotheses that guide the design of our data collection experiments: 

\noindent \textbf{H1: Palette ranking is task dependent.}
We hypothesize that the same categorical palette will exhibit different performance ranks across scatterplot tasks, echoing past heuristic design guidance for color encodings \cite{tominski2008task}.
A palette that performs well for one task will not necessarily remain strong for another task, even when category numbers are held constant.
Tasks such as cluster identification, outlier detection, counting, and distance comparison recruit different perceptual processes and therefore impose different demands on categorical encodings.
Further, the performance gap between strong and weak palettes should vary by task, such that some tasks can be more sensitive to palette quality while others are comparatively robust.
This 
echos the broader task-based literature~\cite{carpendale2008evaluating, saket2018task, sarikaya2017scatterplots}:
visualization effectiveness often depends on the analytical objectives and goals rather than on visual design alone.



\noindent \textbf{H2: Encoding choice moderates task performance.} We anticipate that some tasks may favor shape, whereas others favor color due to differences in their perceptual processing. For example, shape may support tasks requiring segmentation \cite{burlinson2017open} while color may more effectively drive search \cite{frey2008s}.
More generally, the relative utility of color and shape may depend on the perceptual demands of the task and on the limits of categorical perception for each channel.


\subsubsection{Stimulus Generation}
We generated scatterplot stimuli using color and shape encodings to differentiate categories in Experiments~1 and~2. Each scatterplot was rendered as a $400\times400$ pixel chart using D3, with a white background and two orthogonal black axes with 13 unlabeled ticks. Points were rendered using $6\times6$ pixel marks.

\paragraph{Encoding Design.}
We constructed stimuli from a pool of 39 colors and 39 shapes derived from prior work~\cite{tseng2026redundant, tseng2024shape}. The colors were sampled in CIELAB space within the ranges $L^* \in [25, 100]$, $a^* \in [-128, 127]$, and $b^* \in [-128, 127]$, using five evenly spaced levels for $L^*$ and two levels for both $a^*$ and $b^*$~\cite{tseng2026redundant}. Shapes were drawn from common visualization tools~\cite{tseng2026redundant}. For color-only and shape-only conditions, each stimulus used $k$ distinct colors or shapes sampled without replacement from the corresponding pool, where $k$ represents the number of categories in that scatterplot.

\paragraph{Data Generation for Tasks.}
We generated datasets with $k \in \{3, 6, 9\}$ categories to balance experimental coverage and study size. These values reflect commonly used category sizes in prior work~\cite{tseng2026redundant, tseng2024shape} and lie before, at, and after the subitizing bound, allowing us to sample critical performance thresholds found in past studies \cite{wang2025characterizing}. Each dataset was rendered as a scatterplot on a $[0,1] \times [0,1]$ canvas, with point coordinates
between $[0.05, 0.95]$. We generated 50 unique datasets per task and per $k$ value (1,050 total), and applied jittering to all points to avoid overlap. Each category had 20 points in all datasets except for the numerosity estimation task.

Each task applied distinct distributional constraints on the target (answer) category relative to other non-answer categories (see SM for details on the data generation). For example, for \textbf{$T_{cluster}$}, the target category has the smallest convex hull, controlled by Gaussian spread ($\sigma = 0.04$). All stimuli had a 25-50\% difference in convex hull area between the tightest category and all other categories. For \textbf{$T_{spread}$}, the design was inverted: the target category used a larger spread ($\sigma = 0.14$), while other categories were generated from $\sigma \in [0.04, 0.07]$, with a 25--50\% difference in convex hull area to ensure the proper differentiable level between the target and the rest of categories.  Stimulus generation followed task-specific Scagnostics metrics~\cite{wilkinson2005graph} when applicable: \textbf{$T_{cluster}$} used the clumpy metric (tightness via minimum spanning tree~\cite{wang2019improving}), \textbf{$T_{spread}$} used the sparse metric (dispersion via convex hull~\cite{wang2019improving}), and \textbf{$T_{outliers}$} used the outlying metric. Tasks without a direct Scagnostics analogue---\textbf{$T_{localization}$}, \textbf{$T_{correlated}$}, \textbf{$T_{points}$}, and \textbf{$T_{closest}$}---used 
geometric constraints tuned through pilot studies to achieve appropriate difficulty levels and avoid confounding strategies.
All task-specific constraints were tuned through pilot studies to achieve appropriate difficulty levels, avoiding both ceiling effects and excessively difficult conditions.

\paragraph{Stimulus Sampling.}
As we computed pairwise accuracy metrics on 39 colors and 39 shapes~\cite{tseng2026redundant}, exhaustively evaluating the accuracy of all $\binom{39}{2} = 741$ pairs per encoding (3), per category number (9), and per task (7) would require an extremely large study size. We therefore adopted a two-stage sampling strategy to efficiently cover the item space while prioritizing comparisons that are most informative for model estimation.

\noindent \textbf{Stage 1: Anchor-based initial sampling} selected 15 anchor items per encoding (color/shape) to seed the initial Bradley--Terry model. Anchors were chosen to cover the full quality range of the color/shape pool. Specifically, items were divided into three equal performance tiers based on legacy mean pairwise accuracy (top, middle, bottom)~\cite{tseng2024shape, tseng2026redundant}, and five items were sampled from each tier. For color, ties within tiers were resolved by maximizing pairwise CIELAB Euclidean distance among anchors to ensure perceptual distinctness. In addition, up to two items per tier were replaced with those with the highest standard error from a previous baseline study~\cite{tseng2024shape, tseng2026redundant}, prioritizing the resolution of pre-existing uncertainty. This design resulted in $\binom{15}{2} = 105$ anchor pairs per encoding, each evaluated across $k \in \{3, 6, 9\}$. While trials prioritized coverage of anchor-item pairs, non-anchor items were included in each palette to ensure all 39 items accumulated sufficient observations.
See SM for full details of balanced trial and palette assignments.

\textbf{Stage 2: Adaptive sampling} extended coverage to the full 741-pair space using a hybrid adaptive sampling strategy. All candidate color or shape pairs were scored  on three criteria, combined as a weighted sum: \emph{gap filling} (40\%), targeting pairs involving non-anchor items or those with high Stage~1 BT uncertainty; \emph{task shift} (40\%), defined as the maximum deviation in relative pairwise strength from the previous baseline~\cite{tseng2024shape, tseng2026redundant} across the seven tasks; 
and \emph{legacy risk} (20\%), capturing pairs with high uncertainty or near-chance accuracy in prior data. The 60 highest-scoring pairs were selected as Stage~2 targets, ensuring that additional data collection focused on portions of high uncertainty and strong task-dependent variation. Trial assignments followed the same structure as Stage~1. See SM for more details.


\subsubsection{Procedure}
Both experiments followed the same procedure, consisting of three phases: (1) informed consent, (2) task instruction and tutorial, and (3) the formal study. Participants first provided informed consent in accordance with our IRB protocol and completed a demographic questionnaire. They were then introduced to their assigned task, along with examples and three tutorial trials which had to be answered correctly before advancing to ensure task comprehension.
In the formal study, each participant performed one of seven target tasks (see \autoref{tab:tasks}) for 63 trials (60 experimental trials and 3 engagement checks), presented in randomized order.
Participants responded using radio buttons corresponding to each of the $k$ colors or shapes in the presented stimulus.
Each trial was time-limited to 20 seconds.
Engagement checks consisted of simplified stimuli with three categories and substantially lower difficulty than the main trials.




\subsubsection{Participants}
We recruited 158 participants for Experiment~1 and 82 for Experiment~2 using MTurk from the US or Canada, requiring a minimum 95\% approval rating.
Participants who failed more than one engagement check were excluded, resulting in final samples of 147 (97 male, 49 female, 1 DNR; age 26--65) for Experiment~1 and 77 (51 male, 25 female, 1 DNR; age 23--65) for Experiment~2. All participants reported normal or corrected-to-normal vision and were compensated \$2.50 for approximately 10 minutes of participation.


\subsubsection{Analysis}
We used accuracy as the primary dependent measure.
To assess variance across tasks, we computed a Kendall $\tau$ correlation matrix across the seven per-task BT strength rankings.
Low $\tau$ between a task pair indicates that palettes effective for one task are not reliably effective for the other, motivating task-specific modeling.
We also computed mean pairwise accuracy 
as employed in previous work~\cite{gramazio2016colorgorical, tseng2024shape} and used a logistic regression model estimated with Generalized Estimating Equations (GEE) to examine the effect of task on palette effectiveness. 

To model performance per task, for each task $t$, we fit a single Bradley--Terry model across all $k$ values, treating $\log(k)$ as a continuous covariate.
Each trial is expanded into all $\binom{k}{2}$ co-occurring pairs, and the probability of a correct trial response given that items $i$ and $j$ co-appear in the palette is modeled as $\text{logit(}\,P(\text{correct} \mid i, j, k, t)) = \beta_i^t + \beta_j^t + \alpha^t \log(k)$, where $\beta_i^t$ is item $i$’s latent perceptual strength under task $t$, and $\alpha^t$ is the task-specific log($k$) scaling coefficient.
Parameters $\beta_i^t$ and $\alpha^t$ are estimated by minimizing the regularized binary cross-entropy loss using gradient descent on the observed trial correctness.
Infrastructure, anonymized data, and analysis are available on \href{https://osf.io/5r9f6/overview?view_only=c370991e371949d0b8a3a2027f2e51dd}{OSF}.

\subsection{Results}

We first verified that Stage~2 data collection was necessary \fix{using two diagnostics: pair coverage and BT training loss (chance $= \ln(2)\approx 0.693$).
For color, Stage~1 left 232 of 741 pairs (31.3\%) unobserved in each task, but training loss on the observed pairs was already well below chance ($\mu=$ $0.574$, range $0.478$--$0.664$), indicating sufficient observation density per observed pair to separate strong from weak items. Stage~2 reduced unobserved color pairs to 19 (2.6\%, 97.4\% coverage), with only a small additional loss change ($\mu=$ $0.574 \to 0.568$); 
\fix{For color, Stage 2 completed} the comparison graph. For shape, Stage~1 achieved full pair coverage, but training loss remained near chance for all seven tasks ($\mu=$ $0.671$, range $0.646$--$0.689$), indicating that the existing observations were too sparse per pair to reliably separate strong items from weak ones. 
Stage~2
reduced loss substantially by targeting the highest-uncertainty pairs ($\mu=$ $0.671 \to 0.629$; e.g., $T_{spread}$: $0.684 \to 0.566$; $T_{points}$: $0.684 \to 0.567$), 
strategically increasing per-pair density. Stage~2 was thus necessary for both encodings, but for complementary reasons: completing pair coverage for color and increasing per-pair density for shape}.
All analyses below use the combined Experiment~1 + Experiment~2 dataset ($N = 224$).

Overall mean accuracy by task ranged from 73.0\% for $T_{outliers}$ (Most Outliers, 95\% CI [71.0, 75.0]) to 85.4\% for $T_{points}$ (Most Points, [83.8, 87.0]), with the remaining tasks falling between these ranges ($T_{cluster}$: 80.5\% [78.7, 82.2]; $T_{localization}$: 75.9\% [74.0, 77.8]; $T_{spread}$: 84.1\% [82.4, 85.7]; $T_{correlated}$: 77.1\% [75.2, 79.0]; $T_{closest}$: 82.3\% [80.6, 84.0]).
Accuracy also decreased with the category number increased: mean accuracy across all tasks was 93.2\% at $k = 3$ ([92.5, 94.0]), 80.3\% at $k = 6$ ([79.2, 81.5]), and 65.7\% at $k = 9$ ([64.3, 67.1]).
The decline was steepest for $T_{localization}$ (37.3\% difference from $k{=}3$ to $k{=}9$) and $T_{outliers}$ (-36.4\%), and most gradual for $T_{spread}$ (-17.8\%) and $T_{points}$ (-21.3\%), suggesting that tasks relying on spatial extent or numerosity may be more robust to increased category numbers than tasks involving data exploration or outlier detection.

\noindent \textbf{H1: Palette ranking is task dependent.}
To evaluate \textbf{H1}, we computed a Kendall $\tau$ correlation matrix across the seven per-task BT color and shape rankings separately (SM, Fig.~S1).
The $\tau$ values for color are generally low (mean $\tau = 0.12$, median $\tau = 0.11$), and 10 of the 42 unique task pairs are negative, indicating that some color pairs that ranked highly for one task ranked poorly for another.
The strongest reversal is between $T_{cluster}$ (Tightest Cluster) and $T_{localization}$ (Single Half), with $\tau = -0.20$, suggesting that color pairs that worked well for identifying the tightest cluster tended to perform poorly for locating the single-half category, and vice versa.
$T_{cluster}$ also conflicts with $T_{points}$ (Most Points, $\tau = -0.18$) and $T_{outliers}$ (Most Outliers, $\tau = -0.11$), whereas $T_{localization}$, $T_{outliers}$, and $T_{points}$ correlate moderately with one another ($\tau \approx 0.27$--$0.43, p<0.01$), suggesting that color combinations performed comparably across these tasks. 
Shape rankings show a similar pattern of task dependency, though with more modest disagreement (mean $\tau = 0.10$, 6 of 42 pairs negative). See the SM for more detailed shape statistics. 

To characterize task-dependent sensitivity, we divided the tested palettes into three performance groups based on their mean overall accuracy 
and compared mean accuracy across tasks. Palettes in the lowest group has mean accuracy of 54--64\%, the middle 86--94\%, and the highest 97--100\%.
The gap between the lowest and highest performance group varied by task, exceeding 40\% for some tasks and falling closer to 30\% for others, indicating that some tasks are more sensitive to palette quality than others.
These results together support \textbf{H1}: palette performance 
varies meaningfully across tasks.

\noindent \textbf{H2: Encoding choice moderates task performance.}
Color encodings outperformed shape encodings on average (color: $\mu = 82.1\%$, [81.2, 83.0]; shape: $\mu = 77.4\%$, [76.4, 78.4]), but the magnitude of this advantage varied considerably across tasks (SM, Fig.~S2).
The color advantage was largest for $T_{outliers}$ (Most Outliers: $+8.3$\%) and $T_{localization}$ (Single Half: $+6.7$\%), where colors appear to more strongly support search and grouping, and smallest for $T_{closest}$ (Closest to X: $+0.4$\%) and $T_{correlated}$ (Most Correlated: $+1.7$\%), where a heavier reliance on spatial reasoning may lead to smaller differences between encoding channels.
We found a significant interaction effect of task $\times$ encoding 
($\chi^2(6) = 82.88$, $p < .001$), supporting \textbf{H2}: the relative utility of color versus shape is task-dependent. While color consistently outperforms shape overall, the size of this advantage is task-dependent. The gap is trivial for some tasks ($T_{correlated}$, $T_{closest}$: $\leq 2$\%), making shape a practical alternative when color is unavailable, but more substantial for other tasks ($T_{localization}$, $T_{outliers}$: 7--8\%).

\paragraph{Validation of BT estimates.}
We assessed the validity of the per-task BT estimates on two dimensions. For predictive validity, a palette's mean BT $\beta$ score (the average strength of its items) modestly but consistently predicted trial-level correctness across all seven tasks for both encodings (color: $r = 0.12$--$0.24$; shape: $r = 0.16$--$0.24$; all $p < .001$), confirming that BT-derived strength estimates carry a reliable cue. The modest effect compared to mean pairwise accuracy is expected given BT $\beta$ score was constructed in latent space.
For ranking stability, we computed Kendall $\tau$ between the Stage~1-only and combined (Stage~1 + Stage~2) rankings within each task, measuring whether adding Stage~2 data preserved or disrupted the ordering established in Stage~1. For color, within-task $\tau$ ranged from 0.54 to 0.66 across tasks (mean $\tau = 0.60$), indicating moderate-to-substantial convergence that Stage~2 refined rather than reversed the Stage~1 ordering. Shape showed near-zero within-task stability ($\tau$ ranging from $-0.14$ to $+0.14$, mean $\approx 0$), suggesting that the Stage~1 estimates were too noisy to establish a consistent ordering. Therefore, we base our Stage 3 analyses and recommendation system on the combined model.

%% file: sections/sec-05-redundant.tex
Experiments~1 and~2 characterized the perceptual strength of color and shape palettes independently, revealing that item quality rankings vary substantially across tasks. However, visualizations often employ redundant encoding~\cite{nothelfer2017redundant}, where each category is represented using both a unique color and a unique shape. Prior work has shown that redundant encoding can improve accuracy but that color and shape channels interact when used together such that optimizing each channel independently does not necessarily maximize combined accuracy~\cite{tseng2026redundant}. This interaction means that 
a recommendation system must learn \emph{how much} redundancy helps (or hurts) as a function of the task, category count, and the relative quality of each channel.

Rather than exhaustively testing all pairings of 39 colors and 39 shapes---which would be infeasibly large---we systematically sample the redundant encoding space using a $2 \times 2$ quadrant design and measure the \emph{redundant gain} $\Delta G$ across all seven tasks and $k \in \{3, 6, 9\}$. The resulting (quadrant $\times$ task $\times$ $k$) $\Delta G$ values serve directly as training targets for the scoring component of the CatPAL recommendation engine (Section~6), enabling it to estimate the expected benefit of redundant encoding for any given context.

We define the \emph{redundant gain} as $\Delta G = \mathrm{Acc}(\text{color+shape}) - \max\!\bigl(\mathrm{Acc}(\text{color}),\, \mathrm{Acc}(\text{shape})\bigr)$, where $\mathrm{Acc}(\text{color+shape})$ is the observed accuracy under redundant encoding, and $\mathrm{Acc}(\text{color})$ and $\mathrm{Acc}(\text{shape})$ are the empirical pairwise accuracy baselines for each single channel, computed as the mean proportion of correct trials across all $\binom{k}{2}$ item pairs in the set from the Stage~1+2 single-channel data (Experiments~1 and~2), matched by task and $k$. Positive $\Delta G$ indicates that combining channels improves accuracy beyond the better single channel; negative $\Delta G$ indicates interference or ceiling effects.

\subsection{Experimental Design}
\subsubsection{Task \& Stimulus Generation \& Participants}

Experiment~3 uses the same seven tasks and category counts ($k \in \{3, 6, 9\}$) as Experiments~1 and~2, with the key difference that each trial presents a redundant color+shape encoding. Accuracy under this combined encoding is used to compute $\Delta G$.

\paragraph{Four-quadrant design.}
To systematically cover the quality space of color--shape pairings with a tractable study size, we use a $2 \times 2$ quadrant design based on the combined Stage~1+2 BT strength estimates. Items were ranked by their mean $\beta$ across all seven tasks and partitioned into \emph{strong} (top 13) and \emph{weak} (bottom 13) pools within each encoding, yielding four quadrants that span the full range of channel quality combinations.
Color and shape palettes within each quadrant were sampled independently from their respective quality pools. They were then randomly paired within each trial. The quadrant design is fully crossed with all seven tasks and all three $k$ values.

\paragraph{Stimulus pool.} For each quadrant $\times$ $k$ combination, we generated a pool of 100 unique color + shape sets (1,200 total across 4 quadrants and 3 $k$ values), which was shared across all seven task groups. Each participant completed 5 trials per quadrant per $k$ value ($5 \times 4\ \text{quadrants} \times 3\ k = 60$ trials), presented in randomized order. In total, Stage~3 produced 8,400 trial assignments across 140 participants (20 per task $\times$ 7 tasks).

\paragraph{Procedure \& Participants.} Recruitment, procedure, and trial counts matched Experiments~1 and~2 (see \autoref{sec-exp1-exp2}). We recruited 146 MTurk participants and excluded 6 for failing engagement checks, leaving 140 participants (111 male, 28 female, 1 other; age 23--65).

\subsubsection{Analysis}
We computed the redundant gain $\Delta G$ for each (quadrant $\times$ task $\times$ $k$) combination, yielding an empirical map of how much combining color and shape helps or hurts across the full design space. This per-condition $\Delta G$ surface is the primary outcome of Experiment~3: it provides the training signal used in Section~6 to fit the scoring component of the CatPAL recommendation engine. We additionally fit a mixed-effects logistic regression with quadrant, task, $k$, and their interactions as fixed effects to characterize how $\Delta G$ varies across conditions and confirm that the quadrant design captured meaningful variation in redundancy benefit. All anonymized data and analysis code are available on \href{https://osf.io/5r9f6/overview?view_only=c370991e371949d0b8a3a2027f2e51dd}{OSF}.

\subsection{Results}
Overall accuracy under redundant encoding was $\mu = 83.2\%$ (95\% CI [82.4, 84.0], $N = 8{,}400$ trials), slightly above the single-channel means from Experiments~1 and~2 (color: 82.1\%; shape: 77.4\%). Per-task accuracy ranged from 73.9\% for $T_{outliers}$ (Most Outliers, 95\% CI [71.4, 76.4]) to 87.9\% for $T_{points}$ (Most Points, 95\% CI [86.1, 89.8]), with $T_{cluster}$ (87.5\%), $T_{closest}$ (86.1\%), $T_{correlated}$ (83.7\%), $T_{spread}$ (82.5\%), and $T_{localization}$ (80.6\%) falling in between, broadly aligning the task ordering observed in Experiments~1 and~2. However, raw redundant-encoding accuracy 
\fix{meaningfully differs from} redundant gain. By definition, $\Delta G$ compares the combined encoding against the better of a single-channel baseline for the same task and $k$, making it a stricter criterion than comparison with the overall color or shape average.

The mean redundant gain was $\Delta G = -0.023$ across all conditions (95\% CI [$-0.039$, $-0.006$]; $t(83) = -2.73$, $p = .008$; one-sample $t$-test across 
\fix{task $\times$ quadrant $\times$ $k$}), indicating a small but reliable overall interference effect. On average, the combined encoding did not improve accuracy beyond the better single channel. However, this average masked strong systematic variation across the design space.

The effect of $k$ on $\Delta G$ was significant ($F(2, 81) = 18.39$, $p < .001$). At $k = 3$, redundancy was consistently harmful ($\mu = -0.076$, 95\% CI [$-0.097$, $-0.056$]; $t(27) = -7.27$, $p < .001$), likely because single-channel accuracy was already near ceiling and adding a second channel introduced interference.
At $k = 6$, $\Delta G$ was near neutral ($\mu = -0.018$, $p = .195$). At $k = 9$, redundancy became reliably beneficial ($\mu = +0.027$, 95\% CI [$+0.004$, $+0.050$]; $t(27) = 2.28$, $p = .031$), consistent with prior work showing that redundant encoding helps when category numbers are higher~\cite{tseng2026redundant}. This pattern suggests that the value of redundancy depends strongly on task difficulty: it can hurt for low category counts, has little effect at medium load, and becomes useful once discrimination becomes sufficiently difficult.

Task also had a significant effect on $\Delta G$ ($F(6, 77) = 3.78$, $p = .002$). At the task level, $T_{spread}$ (Most Spread) and $T_{outliers}$ (Most Outliers) showed significant negative $\Delta G$ ($T_{spread}$: $\mu = -0.075$, $t(11) = -5.57$, $p < .001$; $T_{outliers}$: $\mu = -0.071$, $t(11) = -2.99$, $p = .012$), suggesting that these tasks are particularly vulnerable to interference when a second channel is added.
By contrast, $T_{localization}$, $T_{correlated}$, and $T_{closest}$ showed $\Delta G$ near zero and were not significantly different from zero (all $p > .40$), indicating that redundancy neither strongly helps nor hurts for these spatial distribution tasks.


Notably, the quadrant factor did not reach significance ($F(3, 80) = 0.15$, $p = .929$). This is still informative: although the quadrant design provided a principled way to sample the space of color--shape quality combinations, the resulting benefit of redundancy was driven more strongly by task and category number than by the coarse strong/weak grouping alone. The fitted interaction model did not reveal a stronger pattern than these main effects \fix{(see Section~\ref{sec:recommend} for how the resulting $\Delta G$ surface is used in the deployed recommendation engine)}.
Together, these findings show that $\Delta G$ varies systematically across the ($k \times task$) space. \fix{As a result, we include
redundant encoding } as a context-dependent adjustment in the CatPAL recommendation engine rather than 
a uniformly beneficial design choice.

%% file: sections/sec-06-model.tex
We used the data from our experiments to build a task-aware palette recommendation system. 
The model input combines two complementary perspectives on palette quality: task-specific BT strength estimates ($\beta_i^t$) and pairwise accuracy metrics that capture how well palettes perform under specific analytic goals.
We also consider low-level perceptual statistics for colors and shapes, such as CIELAB lightness, chroma, and color difference ($\Delta E$) for color and shape-type composition (e.g., open/closed, filled/unfilled) for shape. These features capture potential perceptual cues that may influence effectiveness. Together, these features form a task-dependent scoring model that predicts the probability of a correct trial response for any candidate palette. For redundant color+shape palettes, the score is further adjusted by the expected redundant gain $\Delta G$ estimated in Stage~3. We embedded this model into \emph{CatPAL}, a 
web-based tool that takes a task and category number as input and returns ranked palette recommendations across encoding channels. We validate our trained predictive model against well-known designer palettes and tools \fix{as preliminary evidence for our approach.} 
All trained models, BT matrices, and Stage~3 lookup tables, and source code for model training, and validation are provided in \href{https://osf.io/5r9f6/overview?view_only=c370991e371949d0b8a3a2027f2e51dd}{OSF}.

\subsection{Model Construction}
\label{sec:model-training}

We trained two separate logistic regression models, one for color and one for shape, each predicting whether a participant performs a task correctly on a given task trial with a certain palette. We use logistic regression 
because the training data are sparse and noisy (binary trial outcomes across a pool of 39 items), and the prediction target is binary. In addition, the model provides interpretable coefficients that help us understand feature contributions. More expressive models like neural networks would risk overfitting given the limited palette-level samples. One training instance is constructed per trial, using palette-level features that summarize the full set of colors or shapes presented. The training data are drawn from the combined Stage~1 + Stage~2 dataset (224 participants), resulting in 40,320 training instances per encoding.

Each instance is described by three groups of features. The first captures palette quality through the task-specific BT estimates: the mean, minimum, and standard deviation of $\beta_i^t$ across items in the palette, reflecting the average strength, weakest item, and variance of the presented set under task~$t$. The second group captures empirical pairwise discriminability: the mean and minimum task-specific pairwise accuracy across all item pairs in the palette, derived from per-task per-$k$ pairwise accuracy matrices (7 tasks $\times$ 3 $k$ values) 
computed from Stages~1 and~2. We applied Bayesian smoothing to the cross-task mean to stabilize sparse pairs. The third group captures low-level perceptual properties that previous works~\cite{tseng2023evaluating, tseng2024shape, tseng2026redundant} have demonstrated 
may influence categorical perception.
For color palettes, these are the mean and standard deviation of CIELAB lightness $L^*$, mean chroma $C^*$, and the minimum and mean pairwise $\Delta E$ (CIELAB, perceptual distance). For shape palettes, they are the proportion of filled, unfilled, and open shapes in the palette, reflecting the shape type composition of the palette. All models also include the category count $k$ and task indicator variables (T2--T7, with T1 as the reference category). At inference time, we compute a palette-level score
$\hat{s}(\mathcal{P}, t, k) = \hat{P}!\bigl(\text{correct} \mid \mathbf{x}(\mathcal{P}, t, k)\bigr)$,
where $\mathbf{x}(\mathcal{P}, t, k)$ represents the feature vector for palette $\mathcal{P}$ under task $t$ and category number $k$.

\subsection{Recommendation Engine}
\label{sec:recommend}

Given a task $t$ and category count $k$, the engine scores candidate palettes using $\hat{s}(\mathcal{P}, t, k)$ and returns the top-$N$ results. The task argument can accept a single task (T1--T7), a uniform average across all seven, or a user-specified weighted mixture (e.g., 40\% T1 + 60\% T3). $k$ can be 
any integer from 2 to 10 and is internally mapped to the nearest supported value (2--4 $\to$ 3, 5--7 $\to$ 6, 8--10 $\to$ 9).
For redundant palettes combining color set $\mathcal{C}$ and shape set $\mathcal{S}$, the combined score is $\hat{s}_{\text{redundant}}(\mathcal{C}, \mathcal{S}, t, k) = \max\!\bigl(\hat{s}(\mathcal{C}, t, k),\, \hat{s}(\mathcal{S}, t, k)\bigr) + \Delta G(q, t, k)$, where $\Delta G(q, t, k)$ is the expected redundant gain from the Stage~3 lookup for quadrant $q \in \{\text{SC+SS},\, \text{SC+WS},\, \text{WC+SS},\, \text{WC+WS}\}$, categorized by mean BT strength of $\mathcal{C}$ and $\mathcal{S}$. \fix{Pairing the highest-scoring color with the highest-scoring shape would always select from SC+SS. To generate diverse recommendations, the engine buckets candidate color and shape sets into the four quadrants, selects the top-scoring color and shape within each, and pairs them into that quadrant's representative redundant palette; $\Delta G$ is looked up at the same (quadrant $\times$ task $\times$ $k$) granularity. Although the quadrant marginals are close (SC+SS = $-0.022$, SC+WS = $-0.018$, WC+SS = $-0.019$, WC+WS = $-0.032$; Section~\ref{sec-exp3}), we retain the per-cell lookup as a no-pooling strategy that preserves within-cell signal without claiming quadrant carries explanatory power beyond task and $k$. Quadrant thus serves two roles: bucketing key for candidate diversity (primary), and fine-grained differentiator in the $\Delta G$ lookup that can re-rank palettes with comparable single-channel scores (secondary).}
\fix{All candidates from the four quadrants then compete globally, and the top-$N$ 
palettes are returned from this joint ranking.}

To support colors outside the 39-color experimental pool, the model leverages data from adjacent pool samples. BT strengths do not transfer reliably to novel colors ($R^2 = -0.86$ for nearest-pool beta imputation) and are set to zero for unseen colors. We then estimate pairwise accuracy for unseen colors by mapping each unseen color to its nearest pool neighbor by $\Delta E$ and looking up the corresponding pairwise accuracy from the full 39-item pairwise matrix. If 
two unseen colors map to the same pool color, such pairs are set with a 0.5 chance level, reflecting the assumption that perceptually near colors 
may not be sufficiently discriminable. When users do not request specific colors, 
palettes are constructed using greedy search starting from a randomly seeded color. Each step samples 400 random candidate colors to add to the palette, filters by a minimum perceptual distance ($\Delta E \geq 12$) from existing palette members, scores each valid candidate color, and adds the highest-scoring color to the palette. 
The engine will generate a diverse pool of candidate palettes using multiple random seeds, and the top-$N$ scoring palettes are returned.
When the user supplies target colors to seed the palette, the model's behavior depends on how the number of seeds: if the provided color count exceeds provided $k$, the engine enumerates all $\binom{n}{k}$ subsets (capped at 2,000 for performance considerations) and returns the top-$N$ by score. If fewer than $k$ colors are provided, the model adds the chosen colors to the palette and uses the greedy search process described above to fill the remaining colors.

\subsection{Model Validation}
\label{sec:model-validation}

We evaluate our recommendation engine (CatPAL) 
across
palette ranking fidelity, calibration, and external benchmarking against established designer palettes and tools.


\paragraph{Palette ranking.}
We first assess whether the model correctly orders palettes by expected accuracy---the core requirement for a recommendation engine. 
We evaluate this by 
examining the correlation between the predicted and observed accuracies for each task $\times k$ combination aggregated across all palettes from the experiments using 
Spearman $\rho$.
Across all 21 combinations (7 tasks $\times$ 3 $k$ values), 
$\mu_\rho$ is 0.49 for color and 0.58 for shape, with 
all combinations having a positive rank correlation (see SM, Table~S1). This confirms that the model reliably recognizes which palettes produce higher task accuracy. 

\paragraph{Calibration.}
We examine whether the model’s predicted probabilities align with observed accuracy by grouping trials into ten equal-width bins of predicted $\hat{P}(\text{correct})$ and comparing them with the corresponding empirical means. Overall, the model is well-calibrated, with a mean absolute calibration error of 0.033 for color and 0.019 for shape (SM, Fig~S3), and predicted scores closely tracking observed accuracy across all tested palettes. The largest deviations appear in the lower-performing range for color (0.10–0.20, $\Delta = -0.09$), where the model slightly underestimates accuracy, and in the mid-performing range for shape (0.50–0.60, $\Delta = +0.08$), where it shows mild overconfidence.
Overall, this suggests that the model’s scores can be interpreted not only as relative rankings, but also as reasonable approximations of the probability of a correct response.

\paragraph{Benchmarking against reference palettes and tools.}
\fix{The above assessments confirmed 
that the scorer produces palette rankings consistent with observed human accuracy ($\rho = 0.49$ for color and $0.58$ for shape) and well-calibrated probability estimates for accuracy (mean absolute error of $0.033$ and $0.019$). We  then \emph{applied} the model to estimate expected accuracy for 20 palettes generated by  CatPAL using random seeding. These recommended palettes include colors that lie outside our 39-color experimental pool and are therefore 
not tested in our trials. The predicted-accuracy advantage of CatPAL recommendations over a conventional baseline therefore reflects an estimated real-world accuracy advantage, conditional on the calibration established above. While future work should formally test the recommendation engine against conventional approaches in a user study, this approach provides preliminary evidence of CatPAL's \textit{overall} recommendation approach against a validated scorer (e.g., the designer palettes likely include colors that the engine's random seeding approach does not consider). The Supplement further deconstructs the palettes across color metrics (SM, Table~3). While these metrics are primarily descriptive rather than evaluative---as we lack formalized evaluative metrics for palette design---we note that CatPAL is the only system among the 11 evaluated that achieves top-3 rankings on $L^*_{\text{std}}$, $C^*_{\text{std}}$, and $\Delta E_{\text{mean}}$ simultaneously at every $k$: the three color-feature factors that prior work~\cite{tseng2026redundant} identified as significantly predictive of categorical palette performance.} 
 

We compare CatPAL recommendations against three classes of baselines.
For fixed designer palettes, we evaluate all $\binom{10}{k}$ subsets of Tableau~10~\cite{tableau}, D3~Category~10~\cite{6064996}, ColorBrewer~Set3~\cite{harrower2003colorbrewer}, ColorBrewer~Paired~\cite{harrower2003colorbrewer}, and Stata~S2~\cite{statagraphics19}, scored using the unseen-color approach in Section \ref{sec:recommend} (BT betas set to zero; pairwise accuracy through nearest-pool lookup by $\Delta E$). For tool-generated palettes, we sample five palettes per $k = 3, 6, 9$ from Colorgorical~\cite{gramazio2016colorgorical} and IWantHue~\cite{iwanthue}, and from three large language models (Gemini, ChatGPT, and Claude), each prompted to produce five 10-color palettes (color swatches can be found in SM Fig~S4). Scores are averaged uniformly across all seven tasks and reported at $k = 3, 6, 9$ as Mean (Best), where best reflects the single highest-scoring palette within each sample and mean reflects expected average performance across all samples (\autoref{tab:color-benchmark}).

CatPAL outperforms all baselines on both mean and best performance at every $k$. The designer palettes show high best-case scores (up to 0.984 at $k=3$) but low mean scores (0.060--0.208), 
suggesting that $k$-subsets of these palettes should be selected with care. Colorgorical~\cite{gramazio2016colorgorical} achieves the strongest mean among tool-generated baselines (0.623--0.695 at $k=3$--$6$), while LLM-generated palettes generally have lower scores at small $k$ and show inconsistent improvement at larger $k$. CatPAL achieves mean predicted accuracy of 0.946, 0.970, and 0.891 at $k=3$, $6$, and $9$ respectively, with best-of-20 scores of 0.997, 0.989, and 0.963, substantially above any comparison baseline.

For shape encodings, CatPAL (using a 39 shape pool) is benchmarked against the default shape palettes from D3, Tableau, MATLAB, Excel, and R. CatPAL's top-1\% pool recommendations achieve mean predicted accuracy of 1.000, 0.999, and 0.986 at $k = 3$, $6$, and $9$, respectively. The strongest fixed shape library (MATLAB) reaches best-subset scores of 1.000, 0.996, and 0.780, but with lower mean scores (0.685, 0.950, and 0.710), indicating that only a small subset from the MATLAB shape palette is robust. CatPAL reliably finds those near-optimal palettes, while all other libraries require exhaustive enumeration to identify them. 
Full results can be found in the ``Shape Encoding Benchmark'' section of the Supplement.

\begin{table}[t]
\centering
\small
\caption{Predicted discriminability (Mean (Best)) for color palettes at $k = 3, 6, 9$, averaged across all seven tasks and scored via the unseen-color path. For fixed palettes, mean and best are taken over all $\binom{10}{k}$ subsets; for tool-generated palettes and LLMs, over five independently sampled palettes; for CatPAL, over 20 greedy-search runs.}
\vspace{-0.5em}
\label{tab:color-benchmark}
\begin{tabular}{llccc}
\toprule
\textbf{System} & \textbf{Type} & $k = 3$ & $k = 6$ & $k = 9$ \\
\midrule
\multicolumn{5}{l}{\textit{Fixed designer palettes}} \\[2pt]
Tableau 10          & Fixed & .064 (.984) & .703 (.984) & .471 (.512) \\
D3 Category 10      & Fixed & .093 (.983) & .518 (.936) & .735 (.793) \\
ColorBrewer Set3    & Fixed & .060 (.983) & .676 (.941) & .275 (.360) \\
ColorBrewer Paired  & Fixed & .127 (.983) & .659 (.965) & .651 (.836) \\
Stata S2            & Fixed & .208 (.980) & .729 (.929) & .476 (.528) \\
\midrule
\multicolumn{5}{l}{\textit{Tool-generated palettes}} \\[2pt]
Colorgorical        & Tool  & .623 (.986) & .695 (.866) & .655 (.725) \\
IWantHue            & Tool  & .043 (.211) & .580 (.689) & .452 (.740) \\
Gemini (Flash)        & LLM   & .099 (.273) & .498 (.772) & .394 (.647) \\
ChatGPT (GPT-5.3)       & LLM   & .057 (.273) & .435 (.708) & .521 (.752) \\
Claude (Sonnet 4.6)        & LLM   & .065 (.257) & .676 (.768) & .492 (.743) \\
\midrule
\multicolumn{5}{l}{\textit{CatPAL (model-optimized)}} \\[2pt]
\textbf{CatPAL}     & Opt.  & \textbf{.946 (.997)} & \textbf{.970 (.989)} & \textbf{.891 (.963)} \\
\bottomrule
\end{tabular}
\vspace{-1.5em}
\end{table}

\subsection{The CatPAL System}
\label{sec:catpal}

CatPAL is a public web-based tool that transforms the trained model into an interactive palette design workflow, available at \url{https://catpal-palette-automation.web.app/}.
The interface contains two main sections: a left settings panel for specifying the design context, and a right section for generated and ranked palettes, with additional components for interactive refinement within each palette card. \autoref{fig:teaser} shows a representative interface.

\paragraph{User inputs and task specification.}
Users begin by defining their visualization context with the option to preview palettes either on user-provided (uploaded as a CSV) or random data. \fix{While CatPAL is designed for scatterplots, users can also preview the palettes on bar charts and line graphs to qualitatively assess their applicability to other chart types.} 
Users first select what kind of categorical palette they want---color, shape, or redundant (color-and-shape)---and the category number $k$. 
They can select any number of target tasks from our tested seven (cluster identification, spatial distribution, dispersion, outlier detection, correlation, quantity comparison, proximity judgment), including no tasks, which weights the input from each task model equally. 
When multiple tasks are selected, users may optionally assign weights to reflect the relative importance of each task
to optimize for task-weighted predicted accuracy.

\paragraph{Constraint-based recommendations.}
CatPAL can incorporate user-specified 
colors or shapes 
into the generated recommendations. For shape, the interface provides a picker with 39 shapes and filtering options (e.g., shape type and source library), enabling users to either pin specific shapes or define a custom shape pool. 
The color picker allows users to specify preferred colors from any color in the conventional RGB gamut. 
As described in \autoref{sec:recommend}, any specified colors or shapes are treated as fixed elements of the palette, and CatPAL fills the remaining slots by selecting combinations that maximize predicted accuracy under these constraints. If the requested number of categories $k$ is smaller than the number of user-selected elements, CatPAL instead ranks their subsets.
This design allows users to work within existing brand palettes, libraries, or design systems while still benefiting from empirically grounded recommendations.

\paragraph{Recommended palettes and live preview.}
For color and redundant palettes, CatPAL allows users to control both the diversity and number of recommendations. Users can specify the number of runs, where each run is initialized with a different random seed; increasing the number of runs results in more diverse palettes, at the cost of longer computation time. Users can also set how many results CatPAL returns ($n \in [3-20]$). 
Once generated, CatPAL returns the top-$n$ palettes that satisfy the specified constraints sorted by predicted accuracy. Palettes and their corresponding accuracy are displayed on each palette card. Users can then select any palette card to preview it in a scatterplot, either using their own data or a randomly generated dataset.

\paragraph{Interactive refinement and export.}
Users can leverage the recommendation to interactively edit palettes. Clicking any palette member opens a swap panel showing model-ranked alternatives, each annotated with its predicted accuracy delta relative to the current item (positive deltas in green, negative in red), enabling perceptually-informed editing. 
Colors can be exported in JSON or CSS format as
HEX, RGB, or CIELAB colors. 
A PNG export captures the current scatterplot preview for use in presentations or documentation.


%% file: sections/sec-07-discussion.tex
We introduce an empirically-grounded approach for generating categorical palettes tailored to different analytical tasks. We achieve this by conducting a series of experiments that model palette design parameters across seven common scatterplot tasks, training a probabilistic model on the resulting data, and using this model to drive a novel recommendation engine and palette design system for creating color, shape, and redundant palettes. 
This work offers insight into task-driven design and the potential interplay between empirical research and system design. 

\subsection{Task-Based Effectiveness in Categorical Visualization}

We found that categorical palette effectiveness is strongly task-dependent.
Rather than searching for a universally optimal palette, designers and automated systems should consider task-specific palette optimization.
This finding supports 
the general principle in visualization that design effectiveness 
deeply depends on the user's analytical tasks~\cite{saket2018task,shneiderman2003eyes}.
While prior palette design tools 
optimize for general goals, such as perceptual distinctiveness~\cite{salvi2024color} or aesthetic preference~\cite{gramazio2016colorgorical}, our results demonstrate that task is a key determinant of palette efficacy: palettes that perform well for one task may not perform equally well for another.

The perceptual mechanisms at play in different tasks may cause these differences in performance. For example,
palettes that performed well on outlier detection were also generally effective in sparsity/spread-related tasks. 
Both tasks require people to perceive the spatial dispersion of points and identify those that deviate from main regions, which all heavily rely on accurately estimating the convex hull 
to isolate the impact of specific points~\cite{wilkinson2005graph, wang2019improving}.
In contrast, a different set of palettes best supported tasks like correlation and clustering. 
These tasks likely rely more heavily on mechanisms 
for assessing patterns descriptive of the majority of the category's distribution. 
For example, correlation judgments require perceiving the overall orientation and coherence of a point set, while clustering requires perceiving group structure and density.
These tasks likely rely more on processes such as grouping~\cite{gramazio2014relation} and ensemble perception~\cite{szafir2016four} rather than isolating individual points. 
As a result, palettes that support strong global grouping and pattern perception tend to perform similarly across these tasks but differ from palettes optimized for outlier or sparsity detection.

Our results suggest that scatterplot tasks can be broadly grouped into at least two perceptual categories for palette design: local element detection and global structure perception.
These divisions point to the potential for mechanistically-driven design: future work could enumerate how design can best support different perceptual processes to generalize design choices to a broader and more complex set of tasks. 

\subsection{Empirical Model for Palette Design}

Most prior palette generation tools rely on metric-based optimization.
These methods typically optimize objective functions such as maximizing color distance in perceptual color spaces~\cite{gramazio2016colorgorical} or discriminability metrics~\cite{lu2020palettailor}.
While these metrics are useful proxies for perceptual separability, they do not directly measure how well palettes support actual visualization tasks. They instead offer a grounded formalization of design heuristics.
As a result, metric-based palettes may be perceptually distinct but not necessarily optimal for specific analytical tasks such as clustering, correlation judgment, or outlier detection.

Our approach considers a more top-down approach to palette automation, using 
an empirical and task-driven alternative for metric-based methods.
We directly measure human performance on visualization tasks and model palette effectiveness using 
metrics derived from the corresponding data.
While this approach does not necessarily allow for precise claims about low-level features of design (e.g., nameability or discriminability), it allows CatPAL to consider a more holistic view of design, capturing the rich, intertwined space of perceptual, cognitive, analytical, and environmental factors that affect visualization use in practice. 
Future work should advance palette design towards models that can leverage the strength of both this top-down approach with the bottom-up, metric-based approach used in past work to integrate additional design considerations such as semantics \cite{mukherjee2021context}, aesthetics \cite{gramazio2016colorgorical}, accessibility \cite{salvi2024color}, and affect \cite{braun2025affective}. 

\subsection{Limitations and Future Work}
Our results do not exhaustively test all palettes as doing so would be infeasible due to the exponential number of possible designs. We anticipate that our approach of sampling and modeling palette design using pairwise comparisons 
provides a scalable and empirically grounded method for estimating perceptual performance across large design spaces without exhaustively testing all possible configurations.
Future work should explore the utility of this approach in other design spaces, such as \fix{Gestalt principles~\cite{gramazio2014relation} and ensemble summarization~\cite{szafir2016four}}, to understand the trade-offs in different sampling and modeling techniques for generating empirical design data. 

\fix{While our study focuses on scatterplots, the color palettes we identify likely generalize to chart types with larger colored areas (e.g., bar charts, maps, network visualizations) as the small (6$\times$6 pixel) marks used here represent a conservative test of color discriminability \cite{szafir2018modeling}. The task-dependent rankings, however, are scatterplot-specific. Future work should empirically validate the generalizability of our findings across other chart families.} Encoding effectiveness may also depend on
other factors of the display, such as the background color or assignment of colors to categories. 
Future work should investigate how such factors 
interact with encoding choice and task.

Although our model is trained on a relatively large and diverse set of 39 commonly used shapes, \fix{the fixed library 
means our system cannot score arbitrary shape geometries}. Learning perceptual features directly from shape geometry, rather than relying on a fixed, enumerated set, would enable generalization to novel shapes.
However, achieving this goal may require defining the set of shape features that affect palette design, which remains an open research problem \cite{tseng2024shape}. \fix{This challenge also limits CatPAL's preliminary evaluation as we lack external criteria for assessing shape palettes and existing metrics for color are insufficient to assess overall palette performance \cite{tseng2023evaluating,tseng2026redundant}. Future work should more formally evaluate CatPAL's performance across a wider range of criteria; however, while past work has established measures for assessing continuous colormap efficiency \cite{bujack2017good}, generating evaluative metrics for assessing categorical palette performance remains key future work. }

%% file: sections/sec-08-conclusion.tex
We investigate categorical palette design 
across three crowdsourced experiments. 
We use the results of these studies to compute 
task-specific BT strength estimates and combine these estimates with 
additional perceptual features relevant to palette design to create a 
model for predicting palette effectiveness across tasks. We embedded this model in \emph{CatPAL}, a task-aware system for categorical palette recommendation grounded in empirical perceptual data.
Our results demonstrate that categorical palette design should move beyond task-agnostic perceptual metrics toward task-conditioned, empirically-grounded recommendations that more holistically consider the needs of a given design problem.
\emph{CatPAL} serves 
as an example of how experiments can both advance theory and enable adaptive, data-driven encoding recommendation tools, making empirical data actionable for both theory and practice.